\documentclass[twocolumn,showpacs,superscriptaddress,eqsecnum]{revtex4}
\usepackage{xspace,psfig,epsfig,amsfonts,amssymb}

\newcommand{\beq}{\begin{equation}}
\newcommand{\eeq}{\end{equation}}
\newcommand{\bqa}{\begin{eqnarray}}
\newcommand{\eqa}{\end{eqnarray}}

\newcommand{\erf}[1]{Eq.~(\ref{#1})}

\newcommand{\bra}[1]{\langle{#1}|}
\newcommand{\ket}[1]{|{#1}\rangle}

\newcommand{\sq}[1]{\left[ {#1} \right]}
\newcommand{\cu}[1]{\left\{ {#1} \right\}}
\newcommand{\ro}[1]{\left( {#1} \right)}

\newcommand{\7}{\!\!&\!\!}%{O&O}%
\newcommand{\m}{\;(\!\!\!>}% \triangleright\cdot\!\!

\newcommand{\ZZ}{{\mathbb Z}}
\newcommand{\mod}{\textrm{ mod }}

\newcommand{\col}{0.45\textwidth}
\newcommand{\p}{\!-\!\!\!-\!}
\newcommand{\thp}{\!-\!\!\!-\!\!\!-\!}
\newcommand{\hp}{\!-\!}
\newcommand{\pp}{\!=\!\!=\!}
\newcommand{\x}{\!\times\!}
\newcommand{\cond}{\hp\!\bullet\!\hp}
\newcommand{\til}{\;\cdots\;}

\begin{document}

\title{ROM-based Quantum Computation: Experimental	Explorations  	
using \\ Nuclear Magnetic Resonance, and Future Prospects}
\author{D. R. Sypher}
\affiliation{Department	of Physics,	University of Queensland, Brisbane 4072	Australia.}
\affiliation{Centre	for	Quantum	Dynamics, School of	Science, Griffith University,
Brisbane 4111, Australia.}
\affiliation{Centre	for	Quantum	Computer Technology, University	of Queensland,
Brisbane 4072 Australia.}
\author{I. M. Brereton}
\affiliation{Centre	for	Magnetic Resonance, University	of Queensland,
Brisbane 4072 Australia.}
\author{H. M. Wiseman}
\email{H.Wiseman@gu.edu.au}
\affiliation{Centre	for	Quantum	Dynamics, School of	Science, Griffith University,
Brisbane 4111, Australia.}
\affiliation{Department	of Physics,	University of Queensland, Brisbane 4072
Australia.}
\author{B. L. Hollis}
\affiliation{Department	of Physics,	University of Queensland, Brisbane 4072
Australia.}
\author{B. C. Travaglione}
\affiliation{Centre	for	Quantum	Computer Technology, University	of Queensland,
Brisbane 4072 Australia.}

\begin{abstract}
ROM-based quantum computation (QC) is an alternative to oracle-based 
QC. It has the advantages of being less ``magical'', and being more 
suited to implementing space-efficient computation (i.e. computation 
using the minimum number of writable qubits). Here we consider a number 
of small (one and two-qubit) quantum algorithms 
illustrating different aspects of ROM-based QC. They are: (a) a 
one-qubit algorithm to solve the Deutsch problem; (b) a one-qubit 
binary multiplication algorithm; (c) a two-qubit controlled binary 
multiplication algorithm; and (d) a two-qubit ROM-based version of the 
Deutsch-Jozsa algorithm.  For each 
algorithm we present experimental verification using NMR ensemble 
QC. The average fidelities for the 
implementation were in the ranges 0.9 -- 0.97 for the 
one-qubit algorithms,  and 0.84 -- 0.94 for the two-qubit algorithms. 
We conclude with a discussion of future prospects for ROM-based 
quantum computation. We propose a four-qubit 
algorithm, using Grover's iterate, for solving a miniature 
``real-world'' problem relating to the lengths of paths in a network.
\end{abstract}


\maketitle

%\begin{multicols}{2}
%\narrowtext

\section{Introduction}

The current excitement in, and perhaps even the existence of, 
the field of quantum computation \cite{NieChu00} is due to the 
demonstration that 
quantum computers can solve problems in fewer steps than classical 
computers \cite{Deu85,DeuJoz92,Sho94,Gro96,Gro97}. An  
improvement is rigorously 
established for the Deutsch-Josza algorithm \cite{DeuJoz92} and for 
Grover's search algorithm \cite{Gro97}, while Shor's factorization 
algorithm \cite{Sho94} 
uses exponentially fewer steps than any {\em known} classical algorithm.

It is interesting that, of the above quantum algorithms, those that are provably 
faster (a) are not exponentially faster, and (b) make use of an oracle.
An oracle is a 
``black-box'' that defines a function 
\beq \label{funcdef}
f: \ZZ_{2^{n}} \mapsto \ZZ_{2^{m}}.
\eeq
Here $\ZZ_{N}$  is the natural numbers modulo $N$, that 
is, $\{0,1,\ldots,N-1\}$.
The oracle ${O}_{f}$ acts on a $n$-qubit string  $\ket{k}$, 
and an $m$-qubit string 
 $\ket{l}$ as follows:
\beq \label{oracledef}
O_{f}\ket{k}\ket{l} = \ket{k}\ket{l \oplus f(k)},
\eeq
where $\oplus$ represents bit-wise addition modulo 2. Note that we are 
defining an oracle so that it can be applied to classical bit strings 
as well as to qubit strings.

Although the concept of an oracle is very useful in the context of 
complexity theory, they are, as their name suggests, somewhat 
``magical'' in their operation. Thus they may
hide a great deal of computational complexity in one step, and for 
this reason can be considered
 ``unrealistic'' \cite{Pap94}. In a quantum context  
 it has been suggested that  
 counting oracle calls may be a poor way to study the power of  
 algorithms \cite{BenBerBraVaz97}. Finally, it seems to us that 
 oracle-based  computing is best for studying time efficiency, rather 
than space efficiency. 

All of these factors suggest that it is worth exploring an 
alternative basis for computation. In this paper we explore 
quantum computation based on ROM (Read-Only Memory).
In an earlier paper \cite{TraNieWisAmb02} two of us and co-workers 
showed that a ROM-based quantum computer is 
more space-efficient than a ROM-based classical computer. Here space 
efficiency is defined in terms of the number of {\em writable} qubits 
required. In 
particular, one writable qubit is sufficient to compute any binary function 
of an arbitrary number of ROM bits, whereas two writable bits are 
needed to achieve the same. Also, for a particular one-bit function 
(multiplication of all the ROM bits) evidence was found to support the  
conjecture that one qubit can solve the problem in polynomial time, 
whereas {\em three} bits are 
required for the same.

These results indicate that ROM-based computation is ideal for 
demonstrating space- (and possibly time-) efficiency on small scale 
quantum computers. Of course, at the moment 
small scale quantum computers are all 
we have experimentally. For example, in ion traps 
the number of qubits that can be coherently 
controlled is at most  four \cite{Sac00}, 
and in Nuclear Magnetic Resonance (NMR) experiments on ensembles of 
molecules, the number is at most  seven \cite{KniLafMarTse00}. 
In this paper we explore  
the space-efficient quantum algorithms in Ref.~\cite{TraNieWisAmb02}, 
as well as other ROM-based quantum algorithms, in an NMR context. 

The structure of this paper is as follows. In Sec.~II we 
review   ROM-based computation as defined in Ref.~\cite{TraNieWisAmb02}. 
In Sec.~III we present the simplest space-efficient 
one-qubit algorithm (which solves Deutsch's problem). Sec.~IV 
covers the one-qubit ROM-multiplication algorithm of 
Ref.~\cite{TraNieWisAmb02}. In Sec.~V we present a two-qubit version 
of this, the controlled-ROM-multiplication, which is also provably 
more space efficient than any classical algorithm 
(which would require three bits). In Sec.~VI we explore 
the Deutsch-Josza algorithm using ROM rather than an oracle. In each 
of the sections III--VI we present experimental results following the 
theory, and we discuss these results in Sec.~VII. 
We conclude in Sec.~VIII with a discussion of future prospects for 
ROM-based quantum computation, and in particular we propose a four-qubit 
demonstration of quantum computing solving a ``realistic'' problem (i.e. 
a problem that can be related to the real world and that 
would require more than a second of human thought to 
solve). 

\section{ROM-Based Computation}

We consider quantum computation using qubits, 
 the number of which remains 
fixed throughout the computation.  The 
computer evolves by the operation of {\em gates}, which implement a 
unitary operation on one or more qubits simultaneously. 
It has been 
shown \cite{Bar95} that a single two-qubit gate, such as the 
controlled-NOT gate,
supplemented by all one-qubit gates, is sufficient to perform all 
possible quantum computations in this model. 
%For convenience we will allow for all two-qubit 
%gates since these can be trivially constructed using  controlled-NOT 
%gates and one-qubit gates \cite{Bar95}. 
Unitary gates are of course reversible. 
 This means that in principle the computation can be 
carried out without 
dissipation of information and hence without energy cost 
\cite{NieChu00,Lan61}. 

To make a fair comparison with unitary quantum computation, we must 
consider reversible classical computation. As is well known, 
universal reversible classical computation is not possible with just 
one-bit and two-bit gates. Rather, a three-bit gate such as the 
Toffoli gate or Fredkin gate is required \cite{FreTof82}.
The measurement of the state 
of the qubits (in the computational basis) takes place only at the end of 
the computation. Similarly, initialization (setting a bit to a 
fiducial state such as $\ket{0}$) 
is allowed only at the beginning of the computation. These 
stipulations are necessary to keep the computation non-dissipative.

Before proceeding, let us establish some notation. We 
will write the $n$-bit (or qubit) representation of a number $x \in 
\ZZ_{2^{n}}$ as $\ket{x}$. This is equivalent to the notation 
$\ket{x} = \ket{x_{n-1}}\ket{x_{n-2}}\ldots\ket{x_{1}}\ket{x_{0}}$, 
where $x = \sum_{p} x_{p}2^{p}$. In a `circuit' diagram, the most 
significant bit (MSB), $\ket{x_{n-1}}$, will appear at the 
bottom of the 
diagram, and the least significant bit (LSB), $\ket{x_{0}}$ at the top.

We used this notation already in \erf{oracledef} to specify the action 
of an oracle which implements the function $f$ defined in \erf{funcdef}. 
In ROM-based computation, the function $f$ that is the subject of 
the computation is 
implemented not by an oracle, but by its values $\{f(k):k\in {\mathbb 
Z}_{N}\}$ being stored 
in read-only memory. Specifically, for $f$ as defined in 
\erf{funcdef}, $N\times m$ ROM bits are required to store the function. 
For the simple case $m=1$ (a binary function), we require $N$ bits 
which could be allocated as $f_{0},f_{1}, \cdots ,f_{N-1}$, where 
$f_{k}\equiv f(k)$.
%\beq \label{ketf}
%\ket{f} = \ket{f_{N-1}}\ldots\ket{f_{0}},
%\eeq
%where $f_{k} \equiv f(k)$. 
These ROM bits are not counted in the size 
of the computer. That is to say, the size of the computer is taken to 
be the number of additional (non-ROM) (qu)bits.

To capture the essence of read-only memory, we impose the following 
constraints:
\begin{enumerate}
\item The ROM bits $\{f_{k}:k\}$ can be prepared only in a classical state.
\item For any gate involving the writable qubits, any 
{\em single} ROM bit, $f_{k}$ for some $k$, may act as an {\em additional} 
control bit.
\item No other gates involve the ROM bits.
\end{enumerate}
These three conditions together imply that the ROM bits will always remain 
in the same state. In finite state automata models, space-bounded 
computation can be discussed using Turing machines with two tapes, one of which 
is read-only \cite{Pap94}. This is clearly very similar to the present idea of 
individually-accessed ROM bits. The necessity for placing a constraint 
on the number of ROM bits that can act 
as simultaneous control bits was discussed in 
Ref.~\cite{TraNieWisAmb02}.

The restriction to single-bit ROM access leads to a 
simplification in the representation of ROM in
 circuit diagrams of reversible computation. 
Rather than explicitly using ``wires'' to represent the ROM-bits 
we will simply leave a space at the top of the diagram, 
and write in which ROM bit (if any) is acting as the extra control bit 
for that gate. This suggests an alternative way to conceptualize the 
replacement of the oracle by ROM. An oracle is like an all-knowing 
person who refuses to divulge information except when asked a question 
in a certain way. ROM is like a committee of people who each have one bit 
of information but who refuse to communicate with one another except by 
acting individually upon a device. In this way 
problems in ROM-based quantum computation can be seen to have some  
similarities to 
problems in quantum communication such as in 
Refs.~\cite{BuhCelWig98,BraCleTap99,GalHar01}. 

\section{One qubit Solution to the Deutsch problem}
\label{sec:1qD}
\subsection{Theory}

The smallest quantum computer is obviously one qubit. It turns out that 
this, plus additional ROM bits rather than an oracle, is sufficient 
to solve the Deutsch  problem \cite{Deu85}
for any $n$. The Deutsch problem can be phrased in the following 
way. Given a 
function of the form (\ref{funcdef}) with $N = 2^{n} \geq 4$ and $m=1$, 
find a true statement from the following list:
\begin{itemize}
\item[(A)] $f$ is not balanced.
\item[(B)] $f$ is not constant.
\end{itemize}
A constant function $f$ is one for which $\sum_{k}f(k) = 0$ or $N$; that is, for which
 $f(k) = 0 \; \forall \;k$ or $f(k) = 1 
\;\forall \; k$. A balanced function $f$ is one for which 
$\sum_{k}f(k) = N/2$. Clearly one of (A) and (B) must be true, 
and they both may be true in which case either can be chosen.

Deutsch and Josza found a quantum algorithm that solved this problem using 
$n+1$ qubits and two oracle calls \cite{DeuJoz92}. 
By replacing the oracle with $2^{n}$ ROM 
bits, we 
are able to solve the problem with a single qubit and with one control 
from each ROM bit. If we were concerned with time-efficiency, the 
exponential number of ``ROM calls'' may seem a problem. However here 
we are concerned only with 
space-efficiency.  

The one-qubit algorithm to solve this problem is very simple:
\beq
\begin{array}{cccccccccccccc}
	 \7  \7 f_{0} \7  \7 f_{1} \7  \7 \til \7  \7 f_{N-1} \7 
	 \7 \7 \7 \7\vspace{-0.25ex}\\
	 \7  \7 | \7  \7 | \7  \7  \7  \7 | \7 \7 \7 \7 \7  \vspace{-0.25ex}\\
	\ket{0}\, \7 \p \7 \sq{\frac{2\pi}{N}}_{y}\!\! \7 \p \7 
	\sq{\frac{2\pi}{N}}_{y}\!\! \7 \p \7 \til \7 \p \7 
	\sq{\frac{2\pi}{N}}_{y}\!\! \7 \p \7 \m \7 \pp \7 ~x
\end{array} .\label{cct1}
\eeq
The computer is prepared in the fiducial state $\ket{0}$. 
Each ROM bit, $f_{k} \in \cu{f_{0},f_{1},\cdots ,f_{N-1}}$, 
in turn controls (indicated by the 
vertical line) a rotation on the 
qubit with unitary operator $\sq{\frac{2\pi}{N}}_{y}$. That is, the 
gate is implemented if and only if $f_{k}=1$. Here we are 
using the notation
\beq
\sq{\theta}_{\alpha} = \exp\sq{-i(\theta/2)\sigma_{\alpha}},
\eeq
where $\sigma_{\alpha}$ are the usual $2\times 2$ Pauli matrices, 
with $\alpha \in\cu{ x,y,z}$. For the standard representation of 
these matrices, the basis states are
\beq
\ket{0} = \ro{\begin{array}{c}1\vspace{-0.25ex}\\0\end{array}}\;,\;\;
\ket{1} = \ro{\begin{array}{c}0\vspace{-0.25ex}\\1\end{array}}.
\eeq
 In 
\erf{cct1}, the measurement is 
represented symbolically by an eye: $\;\m\;$, and yields the result 
$x$, a single bit. That this is a classical piece of information is 
represented by the double, rather than single, wire.

If the function is constant, then either it never leaves the state 
$\ket{0}$, or it is rotated by $N\times(2\pi/N)=2\pi$ around the 
${y}$ axis, returning it to the state $\ket{0}$. If the 
function is balanced, it is rotated by $(N/2)\times(2\pi/N)=\pi$ 
around the $y$ axis, putting it into the state $\ket{1}$. If it is 
neither balanced nor constant it will end up in a superposition of 
$\ket{0}$ and $\ket{1}$, so a measurement will yield either result. 
This computation clearly solves the Deutsch problem. If the 
measured state $x$ of the computer is $0$, the answer returned is (B). 
If the measured state is $1$, the answer returned is (A).

To show the superiority of a space-bounded quantum computer over a 
space-bounded classical computer we simply have to prove that a 
one-bit classical 
computer cannot solve the DJ problem. Consider the simplest case, 
where $n=2$, so that $f$ maps $\cu{0,1,2,3}$ to $\cu{0,1}$. 
Since the only possible one-bit gate is a NOT gate $[{\rm N}]$, which 
obeys $[{\rm N}]^{2}=1$, the only one bit operation for this problem is
\beq
[{\rm N}]^{f_{0}p_{0}+f_{1}p_{1}+f_{2}p_{2}+f_{3}p_{3}},
\eeq
 where each $p_{k} \in \cu{0,1}$. Acting on the initial state $0$, this 
 computes the functional $\sum p_{k}f_{k}$ modulo 2. 
 It is trivial to prove that this functional does not distinguish 
 between balanced and constant functions  for any choice of 
 $p_{0},p_{1},p_{2},p_{3}$.

\subsection{Method}

The sample used for all of the following experimental demonstrations
was a 0.1M solution of heavy chloroform, $^{13}$C$^1$HCl$_3$,
dissolved in d6-acetone, CD$_3$COCD$_3$ (for locking
purposes).  Chlorine isotopes $^{35}$Cl and $^{37}$Cl have large
quadrupole moments ($I=3/2$), resulting in extremely short relaxation
times when covalently bonded, on the order of $10\mu s$.
This has the effect of masking scalar coupling between chlorine and
other nuclei~\cite{HarMan78}.  Thus, the chloroform molecule
is effectively a two-spin system, proton and carbon-13, with $I=1/2$
for both spins.  

All spectra were obtained using a \emph{Bruker DRX-500} spectrometer,
for which the magnitude of $H_0$  was approximately $11.6T$.  The
resonance frequencies of the proton peaks were $\nu_{\rm H} = 
500.137849$MHz and $\nu_{\rm C} = 125.77754749$ MHz. 
The scalar coupling was measured to be 
$J = (214.8\pm 0.5)${Hz}. %(Note that Hz = $2\pi$s$^{-1}$.)
Clearly, $J\ll|\nu_{\rm H}-\nu_{\rm C}|$, so that the two spins can be 
resonantly excited independently.  There is more than
$1$kHz separation to the solvent lines, which thus played no part in
the experiment.  All experiments were performed at a temperature of
$(298\pm 0.1)$K. The measured values for the longitudinal $T_{1}$ and 
transverse $T_{2}^{*}$ (including field inhomogeneity effects) relaxation times 
were $T_1(\mathrm H)=(9.7\pm0.2)$s, $T_1(\mathrm 
C)=(11.0\pm0.2)$s, $T_2^*(\mathrm H) = (6.4\pm0.3)$s, and $T_2^*(\mathrm 
C)=(0.2\pm0.01)$s. 
The maximum pulse program time was approximately $20$ms,
significantly less than all of the above values. 

For the one qubit algorithms, the H nucleus was used as it had a far 
narrower linewidth. The initial 
state is the thermal equilibrium state, which has a small excess spin 
in the longitudinal direction (spin up). This pseudo-pure state 
\cite{CorFahHav97} has 
observable signal proportional to that of state $\ket{0}$, as desired.  
A one-qubit gate 
can be implemented by an appropriately phased transverse magnetic field 
pulse (or short sequence of pulses), 
rotating at the resonant (radio) frequency of the nucleus. If a 
particular gate is ROM-controlled then it is implemented only when 
the value of the controlling ROM bit is one. Following the complete 
pulse sequence, a $\pi/2$ transverse pulse is used to shift 
longitudinal spin into the transverse plane, where its precession 
will induce a signal in the RF
coils (the read-out). A positive spectrum indicates an excess of 
spin-up populations before the read-out pulse was applied. The 
presence of the $^{13}$C nucleus, with almost equal population spin up and 
spin down, causes a frequency splitting of $J/2$. Thus the 
observed spectra for the two logical states are of the form
\beq
	\ket{0} :~ \hp\!\!\frac{|}{}\!\!\hp\!\!\frac{|}{}\!\!\hp  ~~,~~~~~~
	\ket{1} :~ \hp\!\!\frac{}{|}\!\!\hp\!\!\frac{}{|}\!\!\hp 
\eeq

The fidelity of the transformation is calculating 
by dividing the area under the spectrum by that which would 
have arisen from a perfect transformation of the thermal signal.
 Since the final readout is equivalent to the average of the results 
 of projective 
measurements in the $\sigma_{z}$ basis of each member of the ensemble, 
the area ratio $R$ can be considered to be due to a mixture of the 
correct result (with probability $F$) and the incorrect result (with 
probability $1-F$), namely $R = F - (1-F)$. Thus the fidelity is 
calculated as $F = (R+1)/2$. 

\subsection{Results}

The Deutsch problem has a deterministic output if one adds the 
promise that the function $f$ is either balanced or constant. In this 
case output (A) indicates that $f$ is constant and (B) that it is 
balanced. With $N=4$, this means that in effect there are only three 
different pulse sequences arising from the algorithm in \erf{cct1}: 
that in which all values of $f$ are zero, that in which two are one, 
and that in which all four are one.

\begin{figure}[htbp]
	\centering
	\includegraphics[angle=-90,width=\col]{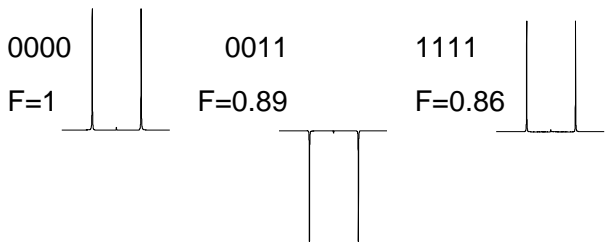}
	\caption{Spectra for H nucleus showing the implementation of our
	one-qubit solution to the Deutsch problem. The values of the four ROM bits 
	are shown above each spectrum. The fidelities (F) are also shown.}
	\label{DJ-1q}
\end{figure}

 The results of these three different pulse sequences are shown in 
Fig.~\ref{DJ-1q}. We see that the results agree well with the theory. 
The first case consists of doing nothing, so its 
fidelity is one, by definition. The fidelity of the other cases is 
calculating as described above. The average fidelity (taking into 
account that there are six possible ways in which the function can 
be balanced) is $\bar{F} = 0.9$.

\section{One qubit multiplication}

\subsection{Theory}

The above results demonstrate that a one-qubit quantum computer 
can solve problems that no one-bit classical computer can. 
By adding one more classical bit, the problem can be solved (this is 
always true, as shown in Ref.~\cite{TraNieWisAmb02}). Moreover,  
there is a two-bit algorithm that is just as time efficient as our 
quantum algorithm above. (It uses the two bits to tally the $f_{k}$s 
modulo 3, and 
is based on the fact that if $2^{n} \mod 3 = 2$ then
$2^{n-1} \mod 3 = 1$ and vice versa.) 

In this section we consider another one-qubit algorithm, which 
is also impossible on a one-bit computer, and which is 
conjectured \cite{TraNieWisAmb02} to be more time efficient than any 
two-bit algorithm. It also solves a more natural problem than the DJ 
problem, namely to find the product of $N$ ROM bits $u_{1},\cdots u_{N}$. 
The quantum 
algorithm 
derived in Ref.~\cite{TraNieWisAmb02} requires exactly $N^{2}$ ROM-calls for 
$N$ a power of two, and $O(N^{2})$ otherwise. The 
required number of ROM-calls $r$ for a two-bit classical computer was found by 
numerical search to be $r=1,3,5,9$ for $N=1,2,3,4$. It is conjectured 
that $r(N)$ is given by the recursion relation 
$r(N) = r(N-1) + 2^{\lfloor N/2 \rfloor}$, and there is an obvious classical 
algorithm requiring exactly this many ROM-calls. This formula is clearly 
asymptotically exponential 
in $n$, but is actually smaller  than the $N^{2}$ ROM-calls in the 
quantum algorithm for $N = 2$, $4$, and $8$. 

In the experiment we only implemented the quantum algorithm for $N=2$ 
and $N=4$. The one qubit algorithm which determines $u_1\x u_2$ can be
constructed as follows.
\beq
\begin{array}{cccccccccccccc}
	 \7  \7 u_1 \7  \7 u_2 \7  \7 u_1 \7  \7 u_2 \7   \7 \7
\7 &\vspace{-0.25ex}\\
	 \7  \7 | \7  \7 | \7  \7 | \7  \7 | \7 \7 \7 \7 & \vspace{-0.25ex}\\
	\ket{0}\, \7 \p \7 \sq{\frac{\pi}{2}}_{y}\!\! \7 \p \7
	\sq{\pi}_{x}\!\! \7 \p \7 \sq{-\frac{\pi}{2}}_{y}\!\! \7 \p \7
	\sq{\pi}_{x}\!\! \7 \p \7 \m \7 \pp & u_1\x u_2
\end{array} .
\eeq
In an abuse of our notation, we will indicate the above algorithm as
\beq
\begin{array}{ccccccc}
	\7 u_1 \x u_2 \7  \vspace{-0.25ex}\\
	\7 | \7  \vspace{-0.25ex}\\
	 \7 \p\sq{\pi}_{y}\!\!\p \7 
\end{array}.
\eeq

Similarly, a gate effecting the transformation
$\sq{\pm\frac{\pi}{4}}_x$, conditional on $u_1\x u_2$, is
\beq
\begin{array}{ccccccccccccc}
	 \7 u_1 \7  \7 u_2 \7  \7 u_1 \7  \7 u_2 
\7  & & \7 u_1 \x u_2 \7\vspace{-0.25ex}\\
	 \7 | \7  \7 | \7  \7 | \7  \7 | \7  & &  \7 |
\7\vspace{-0.25ex}\\
	 \p \7 \sq{\pm\frac{\pi}{4}}_{x}\!\! \7 \p \7
\sq{\pi}_{y}\!\! \7 \p \7 \sq{\mp\frac{\pi}{4}}_{x}\!\! \7 \p \7
\sq{\pi}_{y}\!\! \7 \p & \equiv &  \7 
\p\sq{\pm\frac{\pi}{4}}_x\!\!\p \7 
\end{array}.
\eeq
These operations can be combined to construct an algorithm which
determines the answer $a = u_1\x u_2\x u_3\x u_4$, viz.
\beq
\begin{array}{cccccccccccccc}
	 \7  \7  u_1\x u_2  \7  \7   u_3\x u_4  \7  
\7   u_1\x u_2   \7  \7    u_3\x u_4  \7   \7 \7
\7 &\vspace{-0.25ex}\\
	 \7  \7 | \7  \7 | \7  \7 | \7  \7 | \7 \7 \7 \7 & \vspace{-0.25ex}\\
	\ket{0}\, \7 \p \7 \sq{-\frac{\pi}{4}}_{x}\!\! \7 \p \7 \hp
	\sq{\pi}_{y}\!\!\hp \7 \hp \7 \hp \sq{\frac{\pi}{4}}_{x}\!\!\hp
	 \7\hp\7 \hp\sq{\pi}_{y}\!\!\hp \7 \hp \7 \m \7 \pp
	& a
\end{array}.
\eeq

\subsection{Results}

\begin{figure}[htbp]
	\centering
	\includegraphics[width=\col]{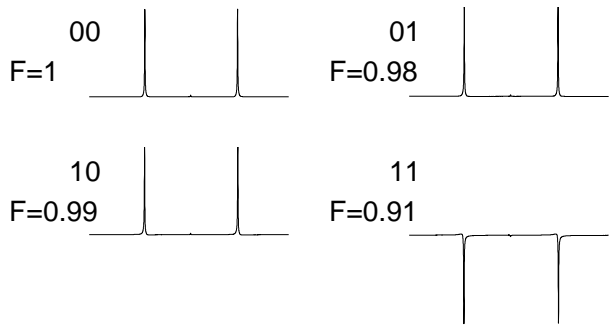}
	\caption{Spectra for the H nucleus for the one-qubit algorithm for 
	multiplying two ROM bits (shown as $u_{1}u_{2}$). A small systematic 
	error is evident in the dispersive features seen in the last case.
	Other details are as in Fig.~1.}
	\label{Mab}
\end{figure}

\begin{figure}[htbp]
	\centering
	\includegraphics[width=\col]{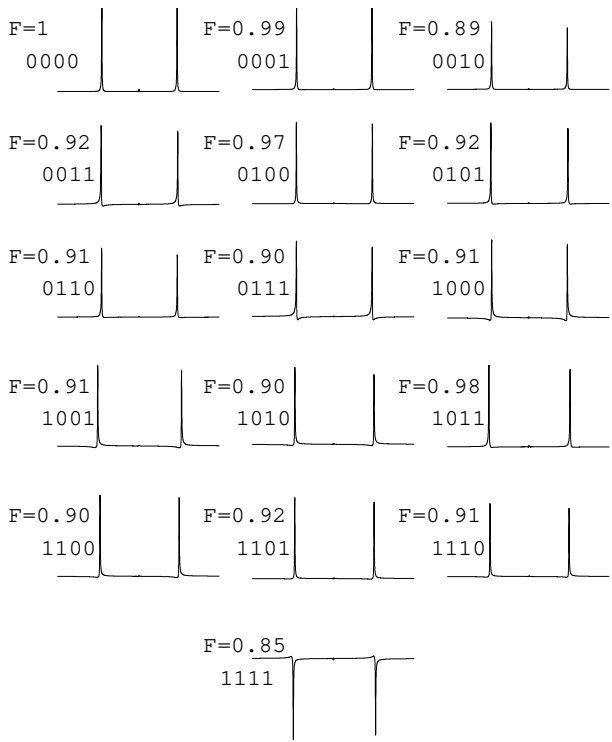}
		\caption{Spectra for the H nucleus for the one-qubit algorithm for 
	multiplying four ROM bits (shown as $u_{1}u_{2}u_{3}u_{4}$). Small systematic 
	errors are evident in the dispersive features seen in most cases.
	Other details are as in Fig.~1.}
	\label{Mabcd}
\end{figure}

The results shown in Figs.~\ref{Mab} and \ref{Mabcd} were obtained by the method outlined above. 
Again we see good agreement with theory. The average fidelity was 
$\bar{F}=0.97$ in the case of multiplying two ROM bits, and 
$\bar{F}=0.92$ for multiplying four ROM bits.

\section{Two qubit controlled-multiplication}

\subsection{Theory}

It is simple to generalize the above single-qubit algorithm (with a 
space advantage of one bit compared to classical computation) 
to a two-qubit algorithm also with a 
space advantage of one bit. This is done by requiring the calculation 
of $u_1\x  u_2 \x  \cdots u_{n} \x  x_{1}$, 
where $x_{1}$ is the value of the second writable bit.  The modified 
algorithm is identical to the one in the preceding section, except that all of the 
gates (which act on the first bit $\ket{x_{0}}$) are controlled by 
the second bit $x_{1}$ as well as by all of the ROM bits. That this 
cannot be done on a two-bit classical computer follows from the 
proof by Toffoli \cite{Tof80} that multi-controlled-NOTs can not be built from
single controlled-NOTs without the use of an auxiliary writable bit.

For the case $n=2$, the circuit is
\beq
\begin{array}{rcccccccccccl}
 \7  \7 u_{1} \7  \7 u_{2} \7 \7 u_{1} \7 \7 u_{2}
\7 \7 \7 &\vspace{-0.25ex}\\
	 \7  \7 | \7 \7 | \7 \7 | \7  \7 | \7 \7 \7 &\vspace{-0.25ex}\\
	\ket{0}\, \7 \p \7 \sq{\frac{\pi}{2}}_{x}\!\! \7 \p \7
\sq{\pi}_{y}\!\! \7\p \7\sq{-\frac{\pi}{2}}_{x}\!\! \7 \p \7
\sq{\pi}_{y}\!\! \7 \p \7
\m \7 \pp  & x_{1}\x u_1\x u_2 \vspace{-0.25ex}\\
\7 \7 | \7 \7 | \7 \7 | \7 \7 | \7 \7 \7  & \vspace{-0.25ex}\\
\ket{x_{1}}\, \7 \p \7 \cond \7 \p \7  \cond \7 \p \7
\hp\cond \hp  \7 \p \7  \cond \7 \p \7 \m \7 \pp & x_{1}
\end{array} .
\eeq
Here the solid circles on the wire for the second qubit indicate that 
it acts as a control qubit for the relevant gate.

\subsection{Method}

As well as selective RF pulses tuned to the H and C nuclei, the
two-qubit algorithm requires an interaction between the nuclei. This 
occurs simply by leaving time between the (negligibly short) pulses 
for the spin-spin coupling Hamiltonian
\beq
H = h J \sigma_{z}^{\rm H}\sigma_{z}^{\rm C}/4
\eeq
to act. These periods of free evolution are usually of duration 
$1/4J$ or $1/2J$, and are simply denoted by this time. For example,
\beq
\sq{\frac{1}{2J}} = \exp[-iH(1/2J)/\hbar]  = \frac{1}{\sqrt{2}}[1 - i\sigma_{z}^{\rm 
H}\sigma_{z}^{\rm C}].
\eeq

The two-qubit algorithm also requires a pure initial state.
A pseudo-pure state $\ket{00}$ of both spins (H and C) up can be prepared 
from the thermal equilibrium state by a sequence of RF pulses, free 
evolution, and gradient pulses. The last of these effectively removes 
the transverse spin of the sample. That is, it  
diagonalizes the state matrix into the logical basis. We use the 
pulse sequence of Cory {\em et al.} \cite{Cor99}, but change the 
$[\pi/6]_{y}$ pulses for both spins into a $[-\pi/6]_{y}$ pulse. This 
is to ensure that the signal is that for the pseudo-pure state 
$\ket{00}$, rather than the negative signal, which corresponds to a 
state matrix $\propto I  - \ket{00}\bra{00}$, where $I$ is the $4\times 4$ identity 
matrix. In theory, this pseudo-pure state preparation procedure 
results in a signal reduction by $3/8$ compared to the original 
thermal equilibrium state. 

The readout is done identically to the one qubit case. 
The correspondence between the logical states and the observed 
spectra is more complicated. The single resonance peak for each spin 
is potentially split into a doublet, at $\omega \pm \pi J$. With the 
NMR convention of frequency increasing from right to left, the 
spectral shapes for the four logical states are as follows.
\beq
\begin{array}{ccc}
	{\rm State} & {\rm Hydrogen} & {\rm Carbon}  \\
	\ket{00} & \thp\!\!\frac{|}{}\!\!\hp & \thp\!\!\frac{|}{}\!\!\hp \\
	\ket{01} & \thp\!\!\frac{}{|}\!\!\hp & \hp\!\!\frac{|}{}\!\!\thp \\
	\ket{10} & \hp\!\!\frac{|}{}\!\!\thp & \thp\!\!\frac{}{|}\!\!\hp \\
	\ket{11} & \hp\!\!\frac{}{|}\!\!\thp & \hp\!\!\frac{}{|}\!\!\thp 
\end{array} \label{table}
\eeq

As in the one qubit cases, for calculating the fidelity we are 
interested only in the occupation of the logical states, since our 
computation is meant to be deterministic. We integrate under the 
spectra at the four frequencies, and divide by 0.375 of the area of the 
original thermal state. The 0.375 is due to the theoretical signal 
loss in the pseudo-pure state preparation described above. This 
procedure gives a number linearly related to the  occupation probabilities 
$p_{00}$, $p_{10}$, $p_{01}$, and $p_{11}$. For example, the 
area ratio $R^{\rm H}_{l}$ under the left Hydrogen peak should 
satisfy
\beq
R^{\rm H}_{l} = p_{10}-p_{11}.
\eeq
In addition, the probabilities should sum to unity. Thus we have five 
equations in four unknowns, which we solve by a least-squares method 
to yield the probabilities. If the desired outcome is $\ket{11}$, for 
example, then the fidelity equals $p_{11}$.

The above algorithm requires four controlled gates. These can be 
constructed from resonant pulses on the two nuclei, plus periods of 
free evolution of duration $1/2J$ or $1/4J$. Specifically, 
\beq
\begin{array}{ccccccccccccccccc}
\p \7 \sq{\frac{\pi}{2}}_{x}\!\! \7 \p & & \p \7
	\sq{\frac{\pi}{2}}_{y}\!\! \7 \p \7
\lceil \7\7 \rceil \7 \p \7 \sq{\frac{\pi}{2}}_{y}\!\! \7 \p \7
\sq{-\frac{\pi}{4}}_{x}\!\! \7 \p \7 \sq{-\pi}_{y}\!\! \7\p\vspace{-0.5ex} \\
\7 | \7 & = & \7 \7 \7 \left|\right.\7 \frac{1}{4J} \7 | \7 \7 \7 \7 \7 \7
\vspace{-0.25ex} \\
\p \7\cond\7 \p & & \p \7 \thp \7 \p \7 \lfloor \7\7 \rfloor
\7 \p \7 \thp \7\p \7\p\thp \7\p \7 \p\p \7 \p
\end{array} \;,
\eeq
\beq
\begin{array}{ccccccccccccccc}
\p \7 \sq{-\frac{\pi}{2}}_{x}\!\! \7 \p & & \p \7
\sq{-\frac{\pi}{2}}_{y}\!\! \7 \p \7
\lceil \7 \7 \rceil \7\p \7 \sq{\frac{\pi}{2}}_{y}\!\! \7 \p \7
\sq{-\frac{\pi}{4}}_{x}\!\! \7 \p \vspace{-0.5ex} \\
\7 | \7 &= & \7 \7 \7 \left|\right. \7 \frac{1}{4J} \7 | \7 \7 \7 \7 \7
\vspace{-0.25ex} \\
\p \7\hp\cond\hp\7 \p & & \p \7 \p\thp \7 \p \7 \lfloor \7\7 \rfloor
\7 \p \7 \thp \7\p \7 \p\thp \7 \p
\end{array} \;,
\eeq
and
\beq
\begin{array}{ccccccccccccc}
\p \7 \sq{\pi}_{y}\!\! \7 \p & & \p \7
\sq{\frac{\pi}{2}}_{y}\!\! \7 \p \7 \lceil \7\7 \rceil \7\p\7
\sq{\frac{2\pi}{3}}_{n}\!\!\7\p\vspace{-0.5ex}\\
\7 | \7 &= & \7 \7 \7 \left| \right.\7 \frac{1}{2J} \7 | 
\7 \7 \7 \vspace{-0.25ex}\\
\p\7\cond\7\p & & \p\7 \thp \7 \p\7 
\lfloor \7\7 \rfloor
\7 \p \7 \p\p \7 \p 
\end{array} ,
\eeq
where $n$ is an axis defined by ${\bf n}=({\bf x}+{\bf 
y})/\sqrt{2}$.

\subsection{Results}

\begin{figure}[htbp]
	\centering
	\includegraphics[width=\col]{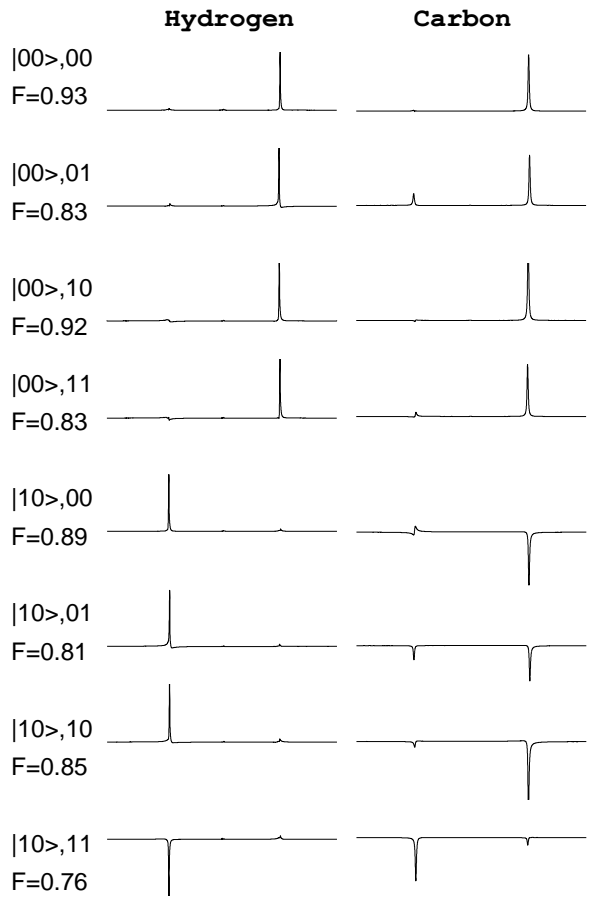}
		\caption{Spectra for the H and C nuclei for the two-qubit algorithm for 
	multiplying two ROM bits, controlled by the first (C) bit. The 
	initial pseudopure states of the two bits are shown with the spectra, 
	along with the ROM bits (shown as $u_{1}u_{2}$). Small systematic 
	errors are evident in most cases.
	Other details are as in Fig.~1.}
	\label{CMab}
\end{figure}

The pseudo-pure state $\ket{00}$ was produced with fidelity of 
$0.93$.  This appears as the first line of Fig.~\ref{CMab}, which 
is a running of the controlled-multiplication algorithm when the ROM 
bits are $00$ (i.e. nothing is done). The results for the other 
possible ROM values appear in the next three spectra. All four of 
these spectra are the same, as they should be since the control 
(Carbon) qubit being set to zero means that $x_{0} = x_{1}\times 
u_{1}\times u_{2} = 0$. The last four 
spectra are repeats of the first four, but with the control (Carbon) qubit 
initially rotated from $\ket{0}$ to $\ket{1}$. This state, 
$\ket{10}$, was prepared with fidelity $0.89$, as shown for the case 
where the ROM bits are $00$. As expected, all spectra but the 
last also have $x_{0}=0$, and the last shows $x_{0}=1$.

The raw fidelities, calculated as discussed above, are 
shown on the figure. Let us scale out the fidelity of the initial state 
preparation (that is, divide the raw fidelities by $0.93$ for the 
first four spectra and $0.89$ for the second four), and take the 
average. Then we get a mean fidelity for the implementation of the 
algorithm of $\bar{F} = 0.94$.

\section{Two qubit ROM-based Deutsch-Josza algorithm}

\label{RBDJA}

\subsection{Theory}

We saw in Sec.~\ref{sec:1qD} that the Deutsch \cite{Deu85} problem can be 
solved on a ROM-based computer with a single qubit. This algorithm was
quite unlike that proposed by Deutsch \cite{Deu85} and Deutsch and 
Josza \cite{DeuJoz92}. In this section we investigate the implementation 
of the Deutsch-Josza algorithm on a ROM-based computer. This requires 
at least two bits to solve the Deutsch problem. Our motivations here 
thus do not include space efficiency. Instead, they are as follows.

First, as noted in the introduction, ROM-based computation seems more 
realistic, so it is interesting to see how it can be applied to an apparently 
quintessentially oracular algorithm.

Second, there is a question of interpretation of past experiments. Again
 as noted in the introduction,  an 
oracle should be definable [see \erf{oracledef}] by its action on a 
classical computer. This is necessary in order not to 
give an unfair space advantage (of $m$ qubits)
to a quantum computer. This requirement is met in  
the original theoretical proposals 
of Deutsch and Josza \cite{DeuJoz92} and Grover \cite{Gro97}. However, 
it is not met in proposals such as that in the 
``refined'' Deutsch-Josza algorithm of Ref.~\cite{ColKimHol98}, 
implemented in Ref.~\cite{Kim99}. That is 
because in this algorithm  the 
oracle directly produces phase shifts, which have no classical analogue.
The requirement of \erf{oracledef} would also rule out 
the oracles implemented in other  NMR experiments 
\cite{JonMosHan98,ChuGerKub98,Van00} (but not to those in 
Refs.~\cite{Chu98,JonMos98,LinBarFre98}). Our analysis here will show that 
these experiments can be very 
easily reinterpreted in terms of ROM calls 
rather than oracle-calls.

Third, there is a question of how quantum the Deutsch-Josza algorithm 
is. The use of a non-classical oracle allows the Deutsch-Josza algorithm to 
be implemented 
using one fewer qubit ($n$ rather than $n+1$). The same number ($n$) of 
qubits are required for the ROM-based implementation. For the minimal case 
$n=2$, it was shown in Ref.~\cite{ColKimHol98} that with the 
non-classical oracle, the Deutsch-Josza algorithm does not utilize 
entanglement. On this basis, the authors claim that it therefore
 ``solves the Deutsch problem in a classical 
way.'' Leaving aside questions as to the meaning of ``classical'' in 
this context, we show that in a ROM-based implementation, 
entanglement necessarily occurs. This suggests that the so-called 
classicality noted in Ref.~\cite{ColKimHol98} is due to the 
unrealistic nature of the oracle they use.
 
In the ROM-based implementation of the Deutsch-Josza algorithm, the 
ROM bits are the same as in Sec.~III, namely 
the binary values $f_{0},\cdots,f_{N-1}$ of the function 
$f$, which is either balanced or constant. For the minimal 
case $N=4$ ($n=2$), the algorithm is
\beq
\begin{array}{ccccccccccccccccccccccccc}
	\7 \7 \7 \7 \7 f_{00} \7 \7 \7 \7  f_{01} \7 \7 \7 \7 f_{10}
 \7 \7 \7 \7 f_{11} \7 \7 \7 \7 \7 \7 & \vspace{-0.25ex}\\
	\7 \7 \7 \7 \7 | \7 \7 \7 \7 | \7 \7 \7 \7 | \7 \7 \7 \7 | \7
 \7 \7 \7 \7 \7 & \vspace{-0.25ex}\\
	\ket{0}\,\7-\7 \sq{\frac{\pi}{2}}_y\!\! \7 \p \7 \lceil\7 \7 \rceil
\7 \p \7 \lceil\7 \7 \rceil \7 \p \7\lceil\7 \7 \rceil \7
 \p \7 \lceil\7 \7 \rceil \7 \p \7 \sq{-\frac{\pi}{2}}_y\!\!\7
 \p \7 \m \7 \pp & x_{0} \vspace{-0.75ex}\\
	\7 \7 \7 \7 \left|\right. \7 \phi_{00} \7 | \7 \7 \left|\right. \7 \phi_{01} 
	\7 \hspace{0.1ex}| \7 
\7 \left|\right. \7 \phi_{10} \7 \hspace{0.1ex}| \7 \7 \left|\right. \7 \phi_{11} \7 | \7 \7 \7 \7 \7
 & \vspace{-0.5ex}\\
		\ket{0}\,\7-\7 \sq{\frac{\pi}{2}}_y\!\!\7 \p \7 \lfloor\7 \7 \rfloor
\7 \p \7 \lfloor\7 \7 \rfloor \7 \p \7\lfloor\7 \7 \rfloor \7
 \p \7 \lfloor\7 \7 \rfloor \7 \p \7 \sq{-\frac{\pi}{2}}_y\!\!\7
 \p \7 \m \7 \pp & x_{1}
\end{array}.
\eeq
Here the four distinct two-qubit gates, $\phi_{\alpha\beta}$
change the sign of the logical state $\ket{\alpha\beta}$, 
leaving the other three unaltered.
Mathematically, the operation of these gates can be expressed as
\beq
\phi_{\alpha\beta}\ket{\gamma\delta} = 
(1-2\delta_{\alpha\gamma}\delta_{\beta\delta})\ket{\gamma\delta}
\eeq
The result $x_{1}x_{0}=00$ indicates a constant function; any other result 
indicates a balanced function.

The four ROM-controlled gates together have exactly the same effect as 
the non-classical oracle introduced in Ref.~\cite{ColKimHol98}. This 
is an example of how any non-classical oracle has a ROM analogue.
The application of an odd number of these gates creates 
an entangled state, so the intermediate states of the quantum 
computer are entangled. It is only because the number of times the 
gates are applied is even (because the function is promised to be 
balanced or constant) that the state at the end of the four 
ROM-controlled gates is not entangled. Entanglement is 
required in all cases of this algorithm except when the function is 
identically zero.

\subsection{Method}

We use the following realization of the two-qubit phase gates:
\beq
\begin{array}{cccccccccccccccccc}
	\p\7 \lceil \7 \7 \rceil\7 \p \7 & & \p \7 \lceil\7\7\rceil\7 \p \7
\sq{-\frac{\pi}{2}}_y \!\! \7 \p \7 \sq{(-1)^{\alpha}\frac{\pi}{2}}_x\!\! \7
\p \7 \sq{\frac{\pi}{2}}_y \7 \p  \vspace{-1ex}\\
	\7 \left|\right.\7\phi_{\alpha\beta}\7\hspace{0.1ex}| \7 \7 & = & \7 
	\left|\right.\7\frac{1}{2J}\7 \hspace{0.1ex}| \7 \7 \7 \7 \7
\7 \7  \vspace{-0.5ex}\\
	\p\7 \lfloor \7 \7 \rfloor\7 \p \7 & & \p \7 \lfloor\7\7\rfloor\7 \p \7
\sq{-\frac{\pi}{2}}_y\!\! \7 \p \7 \sq{(-1)^{\beta}\frac{\pi}{2}}_x\!\! \7
\p \7 \sq{\frac{\pi}{2}}_y\!\! \7 \p
\end{array}.
\eeq
% \rfloor, \rceil, \lfloor, \lceil

\subsection{Results}

\begin{figure}[htbp]
	\centering
	\includegraphics[width=\col]{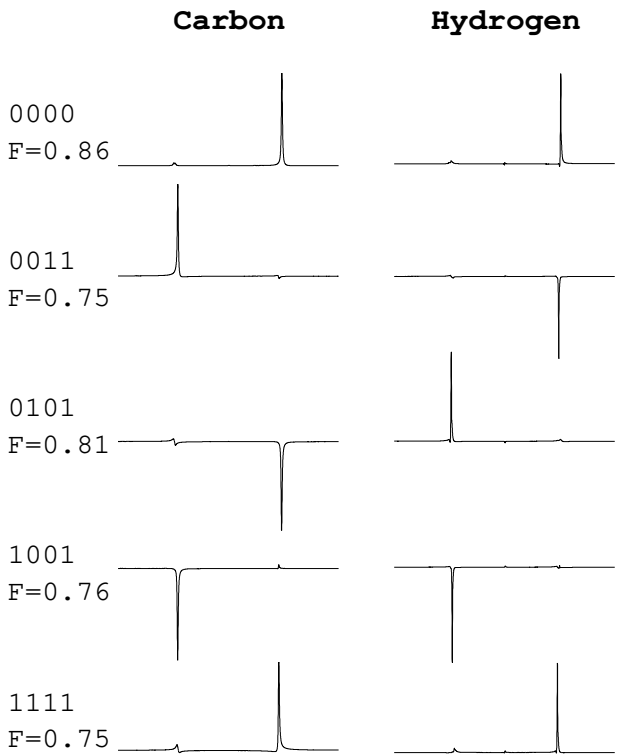}
		\caption{Spectra for the H and C nuclei for the two-qubit ROM-based 
		Deutsch-Josza algorithm. The ROM bits are shown as 
		$f_{00}f_{01}f_{10}f_{11}$. Note that the order of the Carbon and 
		Hydrogen spectra are reversed as compared to Fig.~\ref{CMab} and 
		\erf{table}. Small systematic 
	errors are evident in all cases. Other details are as in Fig.~1.}
	\label{DJ-2q}
\end{figure}

The results are shown in Fig.~\ref{DJ-2q}. Comparing with the 
table (\ref{table}), we see that the result $00$ is obtained only for the case 
of function values $0000$ and $1111$, as expected. Results for the  
values $1100$, $1010$ and $0110$ were not obtained. 
If the algorithm implementation were perfect then the state after the four 
controlled phase gates in these cases would be the same as for 
$0011$, $0101$, and $1001$ shown in Fig.~\ref{DJ-2q}.
There is also no reason 
from the pulse sequence 
to expect the fidelities to be very different for those cases 
implemented. We can 
thus take the cases shown as being representative, and use them to 
calculate an average fidelity over the eight possible functions $f$. 
Scaling away the fidelity of $0.93$ 
for the pseudo-pure state preparation (as in Sec.~V), we obtain
$\bar{F} = 0.84$.

\section{Discussion of Experimental Results}

The experimental results shown clearly verify the ROM-based 
quantum algorithms discussed in this paper. The average fidelities of the 
algorithms were not unusually good for NMR experiments. They 
were $0.9$, $0.97$ and $0.92$ for the one qubit algorithms, and $0.94$ 
and $0.84$ for the two-qubit algorithms. These fidelities are 
definitely correlated with the length of the pulse sequence required. 
However, it is interesting that the 
lowest fidelity (0.84) was 
obtained for the the 
ROM-based Deutsch-Josza algorithm, which was not much longer than the 
other 2-qubit algorithm, but which differed from it in that 
used entangling operations.

The largest source of error was probably spatial and temporal 
variation in the intensity and phase of the
RF pulses.   The spatial variation
of field strength is a direct consequence of limitations imposed by
the structure of the coils, being small Helmholtz coils. 
Cummins and Jones~\cite{CumJon00} provide a good discussion of the
errors incurred in NMR computing and explain how the systematic errors
due to $H_1$ field inhomogeneities can be greatly reduced.  
We did not attempt to apply these techniques.

\section{Future Prospects: A ``Realistic'' Problem}
\label{realprob}
In this paper we have presented a number of ROM-based quantum 
algorithms for one- and two- qubit processors. One of these 
(one qubit multiplication) was derived in Ref.~\cite{TraNieWisAmb02}; the 
rest are new. They include a one-qubit algorithm solving the Deutsch 
problem, a two-qubit controlled-multiplication algorithm, and a two-qubit 
ROM-based Deutsch-Josza algorithm solving the Deutsch problem. 
For all algorithms we have also presented experimental 
verification, using NMR ensemble quantum computing.

All bar one of the above algorithms demonstrated space efficiency, in that 
a classical computer would require an extra processor bit to solve 
the problems. The exception is the ROM-based Deutsch-Josza algorithm. 
We believe that future prospects for ROM-based quantum computation 
lie more in the direction of this last example. There are two 
reasons. First, it follows from the results of 
Ref.~\cite{TraNieWisAmb02} that the maximum space efficiency offered by 
ROM-based quantum computation is one bit, which could not be significant 
in computations of a useful scale. Second, in showing how an 
oracular algorithm  
can be implemented on a ROM-based computer, the example of Sec.~\ref{RBDJA} 
illustrates how time-efficient 
quantum computing could be implemented realistically.

In the remainder of this section we will explore a future prospect for 
ROM-based quantum computation along these lines. We take as our basis 
not the Deutsch-Josza algorithm, but the other famous oracle-based 
algorithm, 
due to Grover \cite{Gro97}. We will show how the oracle in this algorithm 
can also be 
implemented using ROM-calls. We find a specific implementation with two 
properties of interest. First, it  is 
experimentally feasible in the short term, requiring only four 
qubits. Second, it solves a problem that can be related to a read-world 
situation (albeit a miniaturized one), 
and that would require more than a second of 
cerebral processing time to solve.

The problem we consider relates to the lengths of paths between two 
vertices in a network 
(a set of vertices connected by edges of differing lengths). Some 
problems of this nature, such as finding the longest such path, are 
known to be {\bf NP}-complete \cite{Pap94}. This is a class of problems that are 
almost certainly exponentially hard to solve, and are thus of great 
practical interest. 

Consider the network below
\beq
\begin{array}{ccccc}
 \7 \7U\7 \7 \vspace{-0.5ex} \\
 \7 \nearrow \7\uparrow\7 \searrow \7 \vspace{-0.5ex}\\
B\7 \7\hspace{-0.1ex}|\7 \7E\vspace{-0.5ex}\\
 \7 \searrow \7\downarrow\7\nearrow\7\vspace{-0.5ex}\\
 \7 \7L\7 \7
 \end{array}~. \label{network}
\eeq
There are four 
vertices, labeled B, L, U, 
and E (for Begin, Lower, Upper, and End), linked by  
edges. These could represent cities and roads respectively. 
We are interested in the length of paths from B to E, as indicated by 
the direction of the arrows in \erf{network}.  
Assuming no back-tracking, there are four possible paths: 
BLE, BUE, BLUE, and BULE, to which we assign the numbers from 0 to 3. 
 Each path $p$
has a length $L(p)$ associated with it, equal to the sum of the 
lengths $l(e)$ of each 
edge $e$ of the path. Six 
edges must be distinguished, as the edges UL and LU could well 
have different associated lengths. This is because ``length'' 
could represent some generalized cost, such as the time a traveler 
would have to wait for a lift. We discretize the problem by 
assuming that for all $e$, $l(e)$ is either zero or one. Thus for 
all $p$, $L(p) \in \{0,1,2,3\}$. 

The obvious question a traveler would like to ask is, what is the 
shortest path? Unfortunately, this is not the sort of question 
that Grover's search algorithm will answer straight away. 
Rather, what it can answer 
is questions like, what is the path of length 1? If there is 
exactly one path of length 1, Grover's algorithm will find it. If there 
are none or two paths of length 1, Grover's algorithm will return one of 
the four paths at random. If there are three paths of length 1, 
Grover's algorithm 
will actually return the only path that is {\em not} of length 1! Thus, 
what Grover's algorithm really does in our case, is to return a 
number $p$ which means that 
\begin{quote}
path $p$, or none of the paths, is the 
odd-one-out with respect to having the length $L$.
\end{quote}
Here the odd-one-out is the only one having, or the only one lacking, 
a property.

With a little thought it is apparent that actually we are not limited 
to making a demand about a specific length $L$. Rather, we can demand 
information about a set ${\mathbb S}$ of lengths $L$. Since the total 
length $L \in \{0,1,2,3\}$, there are seven such sets, 
\bqa
{\mathbb S}_{1} &=& \{0\}, \\
{\mathbb S}_{2} &=& \{1\}, \\
{\mathbb S}_{3} &=& \{0,1\}, \\
{\mathbb S}_{4} &=& \{2\}, \\
{\mathbb S}_{5} &=& \{0,2\}, \\
{\mathbb S}_{6} &=& \{1,2\}, \\ 
{\mathbb S}_{7} &=& \{0,1,2\}.
\eqa
There are other nontrivial subsets of $\{0,1,2,3\}$, but they are the 
complements of the above seven sets, so they would lead to demands  
already covered by the above seven sets. Specifically, the seven 
demands we could make on our computer are, with $j \in \{1,\cdots,7\}$,
\begin{quote}
What is a path $p$ such that $p$, or none of the paths, is the 
odd-one-out with respect to having a length $L(p) \in {\mathbb 
S}_{j}$?
\end{quote}
In a ROM-based computation, there are six ROM bits to encode the 
lengths
\beq
l_{\rm BL}, l_{\rm BU}, l_{\rm LU}, l_{\rm UL}, l_{\rm LE}, \textrm{ 
and }l_{\rm UE}.
\eeq
and three to encode one of the seven sets ${\mathbb S}_{j}$,
\beq
j_{0},j_{1},\textrm{ and }j_{2},
\eeq
 \\ where these are the bits in the binary representation of $j$.
The values of the nine ROM bits thus code for 
448 different instances of the general 
problem.

Note that the binary representation $j_{2}j_{1}j_{0}$ of $j$ is related to the set 
${\mathbb S}_{j}$ as follows:
\beq
{\mathbb S}_{j} = \{ L: j_{L}=1 \}
\eeq
Using this we can construct a solution to the above problem by the 
following circuit:
\begin{widetext}
\beq
\begin{array}{ccccccccccccccccccccccccccccccccccccccccccccccc}
	\7 \7 \7 \7 \7 l_{\rm BL} \7 \7 \7 \til \7\7 \7 l_{\rm UE}
 \7 \7 \7 \7 j_{0}  \7 \7 \7 \7 j_{1} 
 \7 \7 \7 \7 j_{2} \7 \7 \7 \7 l_{\rm BL} \7 \7 \7 \til \7\7 \7 l_{\rm UE} \7 \7 \7 \7 \7 \7 & 
 \vspace{-0.25ex}\\
	\7 \7 \7 \7 \7 | \7 \7 \7\7 \7 \7 | \7\7\7\7 | \7\7\7\7 | \7 \7 
	\7\7 | \7\7 \7 \7 | \7 \7 \7\7 \7 \7 | \7
 \7 \7 \7 \7 \7 & \vspace{-0.25ex}\\
	\ket{0}\,\7 \p \7 \thp \7 \p \7 \lceil\7 \7 \hspace{-0.05ex}\rceil
\7 \p \7 \til \7 \p\7  \lceil\7 \7 \hspace{-0.05ex}\rceil \7 \p \7\lceil\7\7\rceil \7\p 
\7\lceil\7\7\rceil \7\p\7\lceil\7\7 \rceil \7\p\7\lceil\7 \7 \hspace{-0.1ex}\rceil
\7 \p \7 \til \7 \p\7 \lceil\7 \7 \rceil \7 \p \7 \ket{0} \vspace{-0.5ex}\\
	\7 \7 \7 \7 \left|\right. \7   {}     \7 | \7 \7   \7 \7 \left|\right. 
	\7  {}    \7 | \7 \7 \left|\right. \7 \phi_{00} \7\hspace{0.1ex}|\7 
	\7\left|\right.\7\phi_{01}\7 \hspace{0.05ex}|\7 \7\left|\right.\7\phi_{10}\7\hspace{0.05ex}|\7 
\7 \left|\right. \7   {}     \7 | \7 \7  \7 \7 \left|\right. 
	\7  {}    \7 | \7 \7 \7\7
 & \vspace{-0.5ex}\\
		\ket{0}\,\7\p\7 \thp \7 \p \7 \left|\right.\7 \7 |
\7 \p \7 \til \7 \p \7 \left|\right.\7 \7 | \7 \p \7\lfloor\7\7\rfloor\7 
\p\7\lfloor\7\7\rfloor\7 \p
\7 \lfloor \7 \7\rfloor \7\p \7\left|\right.\7 \7 |
\7 \p \7 \til \7 \p \7 \left|\right.\7 \7 | \7 \p \7 \ket{0} \vspace{-0.75ex} \\
 	\7 \7 \7 \7 \left|\right. \7 U_{\rm BL} \7 | \7 \7   \7  \7 \left|\right. \7 
 	U_{\rm UE} \7 | \7 \7\7\7\7 \7\7\7\7 \7\7\7\7
\7 \left|\right. \7 U^{-1}_{\rm BL} \7 | \7 \7  \7  \7 \left|\right. \7 
 	U^{-1}_{\rm UE} \7 | \7 \7 \7 \7 \7
 & \vspace{-0.5ex}\\
 	\ket{0}\,\7 \p \7 \sq{H} \7 \p \7 \left|\right.\7 \7 |
\7 \p \7 \til \7 \p \7 \left|\right.\7 \7 | \7 \p  \7 \hp \7\thp\7\hp 
\7\p\7 \hp\7 \thp\7 \hp \7\p \7 \hp\7 \thp\7\hp
\7 \p \7 \left|\right. \7 \7| \7\p \7  \til \7 \p \7 \left|\right.\7 \7 | \7 \p \7 
\sq{H}\7 \p  \7 \lceil\7\7\rceil \7
 \p \7 \sq{H}\7 \p \7 \m \7 \pp & p_{0} 
 \vspace{-0.5ex}\\
%\7 \thp \7 \p\7\hp\7\thp\7\hp\7\p 
 	\7 \7 \7 \7 \left|\right. \7   \7 | \7 \7   \7  \7 \left|\right. \7 
 	  \7 | \7 \7\7\7\7 \7\7\7\7 \7\7\7\7
\7 \left|\right. \7   \7 | \7 \7  \7  \7 \left|\right. \7 
 	  \7 | \7 \7 \7 \7 	\left|\right. \7 \phi_{00} \7 \hspace{0.05ex}| \7 \7 \7 
 & \vspace{-0.5ex}\\
		\ket{0}\,\7 \p \7 \sq{H} \7 \p \7 \lfloor\7 \7 \hspace{-0.05ex}\rfloor
\7 \p \7 \til \7 \p \7 \lfloor \7 \7 \hspace{-0.05ex}\rfloor \7 \p  \7 \hp \7\thp\7\hp 
\7\p\7 \hp\7 \thp\7 \hp \7\p \7 \hp\7 \thp\7\hp
\7 \p \7 \lfloor \7 \7\hspace{-0.1ex}\rfloor \7\p \7  \til \7 \p \7 \lfloor\7 \7 
\rfloor \7 \p \7 \sq{H}\7 \p 
\7\lfloor\7 \7 \rfloor \7
 \p \7 \sq{H}\7 \p \7 \m \7 \pp & p_{1} 
 \vspace{-0.5ex}\\
%\7 \thp \7 \p\7\hp\7\thp\7\hp\7\p\7\thp\7\p
\end{array}. \label{4qG}
\eeq
\end{widetext}

The first pair of Hadamard gates \cite{NieChu00} puts the $\ket{0000}$ state into the state 
\beq \label{go}
(1/2)\sum_{p=0}^{3} \ket{p}\ket{00},
\eeq
where the lower two qubits, encoding the path, are in a superposition of 
all four possible paths. The next six gates, conditioned on the six 
ROM bits storing the lengths of the edges, use the upper pair of bits to count the length of each path. 
For example,
\beq \label{getlength}
U_{\rm BL}\ket{p}\ket{L} = \ket{p}\ket{L+\chi[\textrm{BL} 
\sqsubset p]}.
\eeq
Here $\chi[H]$ is the characteristic function, equal to 1 if its 
argument $H$ is true, and zero otherwise, while  
``BL $\sqsubset p$'' means ``BL is an edge in path number $p$''. These functions 
are `hard-wired' in the gate construction, as they depend only of the 
geometry of the network in \erf{network}, not the lengths. The 
addition in \erf{getlength} is defined modulo 4, which is reversible on 2 bits, 
so that this is a well-defined gate. After all 
six such gates have acted, the state of the processor is
\beq \label{superpos}
(1/2)\sum_{p=0}^{3} \ket{p}\ket{L(p)}.
\eeq

Upon this superposition of all paths, and their associated lengths, we 
now act the sign change that is at the heart of Grover's algorithm. 
The three gates controlled by $j=4j_{2}+2j_{1}+j_{0}$ produce the overall 
phase shift
\beq
\ket{p} \ket{L} \to \ro{ 1-2\chi[L\in {\mathbb S}_{j}]}\ket{p}\ket{L}.
\eeq
This changes the sign of the components of the superposition 
(\ref{superpos}) for which the 
length is in the specified set ${\mathbb S}_{j}$. After the application of 
 the inverse of the six controlled gates $U_{\rm BL} \cdots U_{\rm 
UE}$, the processor is in the state
\beq \label{wo}
(1/2)\sum_{p=0}^{3} \cu{1-2\chi[L(p) \in {\mathbb S}_{j}]}\ket{p}\ket{0}.
\eeq

Past experimental implementations of Grover's algorithm 
\cite{JonMosHan98,ChuGerKub98,Van00} have relied upon 
a non-classical oracle that takes the processor directly from a 
state like \erf{go} to one like \erf{wo}. In that case, the upper 
pair of qubits in \erf{4qG} are of course superfluous. For these 
experiments, where there is no real problem being solved, such 
an oracle seems reasonable enough.  However, in the 
context of the present problem, which relates to a ``real-world'' 
situation --- the network (\ref{network}) --- 
an oracle like this would indeed be magical. The point of the 
above analysis is to show  
that, with the help of just two additional qubits, the required oracle 
can nevertheless be implemented in a realistic manner, using ROM-calls.

A Grover iterate is complete with the application of Hadamard gates 
to the lower pair of qubits, and a phase change to the $\ket{00}$ state 
\cite{fnDJ}. 
In this case, because the number of paths is four, a single Grover 
iterate suffices. The final step of the algorithm is to apply the 
Hadamard gates again. After this, the state of the lower pair of qubits of 
the processor is $\ket{p_{j}}$, where $p=p_{j}$ is the 
unique solution of $L(p) \in {\mathbb S}_{j}$ or $L(p) \in \bar{\mathbb 
S}_{j}$, 
where $\bar{\mathbb S}_{j}$ is the complement of ${\mathbb S}_{j}$.
If neither of these equations have a unique solution, then the final state is a 
 superposition of all possible paths, as in \erf{go}. Thus it is 
 apparent that this algorithm does indeed fulfill a demand of the form 
 above.

The above algorithm is certainly not space-efficient. From the results 
of Ref.~\cite{TraNieWisAmb02} it follows that even a classical 
computer could solve this problem with a two-bit processor (although 
probably in more steps). Nor do we claim that it is time-efficient. A 
classical four-bit processor may well be able to solve the problem in 
fewer steps. However it would be interesting to determine what the 
effect would be if one stipulated that the three demand-specifying ROM-bits 
(the bits of $j$) only be used 
once, as in the above quantum algorithm. A similar `once-only' 
constraint on information access was considered in Ref.~\cite{GalHar01}

A lack of both time and space efficiency for this particular 
algorithm would not render it worthless. Consider the 
one-qubit ROM-multiplication algorithm in Sec.~IV. The specific instances 
of that algorithm we discussed and experimentally 
implemented were not time-efficient. However, when scaled up to 
larger numbers of ROM bits, the algorithm was (we conjectured) 
time-efficient compared to the minimal (two-bit) classical processor 
needed to solve this problem. Similarly, for large 
problems, a suitable generalization of this ROM-based Grover algorithm would,  
we hope,  become 
quadratically time-efficient compared to any 
classical algorithm. 

If Grover's algorithm cannot be applied to 
``real-world'' problems in this 
way, then it is of very limited utility. Arguments pointing to the  
generality of its quadratic speed up for large problems have been 
made by two groups. First, Brassard, H\o yer and Tapp  
\cite{BraHoyTap98}
showed how Grover's algorithm can work even if the number of 
``marked'' elements is unknown. Second, Cerf, Grover and Williams 
\cite{CerGroWil00} claimed   
that Grover's algorithm can make use of the structure of a large scale 
problem in the same way as classical search algorithms, and thus 
maintain a quadratic speed up. However, serious doubts on this matter have also been 
expressed \cite{Zal00}, so the question remains open.

Experimentally implementing the above four-qubit algorithm is well 
beyond the 
scope of this study. We have not even compiled it into the appropriate 
``machine language'' (NMR pulse 
sequences). However, this could be done along the same lines as the 
other algorithms presented here, by first decomposing the four-qubit gates 
in \erf{4qG} into sequences of one and two-qubit gates. We hope that the 
challenge of experimental implementation on this, or some similar 
algorithm, is taken up. Although it would still be only a toy 
calculation, it would be a significant step on the way towards full-scale calculations of 
difficult problems pertaining to real-world situations. \\

\acknowledgments
The authors wish to thank S. Crozier, G. Milburn, R. 
Laflamme, E. Knill, A. White, and M. Nielsen for support and advice 
over the course of this investigation.  
HMW also gratefully acknowledges discussions long-passed, but formative,  with G. Toombes 
relating to Sec.~\ref{realprob}. 
This work was supported by the Australian Research Council, the 
University of Queensland, and Griffith University.


\begin{thebibliography}{99}
\bibitem{fnDJ}
 It is interesting to note that if these steps are omitted, 
the algorithm constructed above will still answer certain questions. 
They are questions like that of the Deutsch-Josza 
problem, giving information like ``the number of paths 
having lengths in ${\mathbb S}_{j}$ is not two''.
\end{thebibliography}
\end{document}